# Phase-locked photon-electron interaction without a laser

Masoud Taleb[1,2], Mario Hentschel[3], Kai Rossnagel[1,2,4], Harald Giessen[3], and Nahid Talebi[1,2*]

[1]*Institute of Experimental and Applied Physics, Kiel University, 24098 Kiel, Germany*

[2]*Kiel Nano, Surface and Interface Science KiNSIS, Kiel University, 24118 Kiel, Germany*

[3]*4th Physics Institute and Research Center SCoPE, University of Stuttgart, 70569 Stuttgart, Germany*

[4]*Ruprecht Haensel Laboratory, Deutsches Elektronen-Synchrotron DESY, 22607 Hamburg, Germany*

E-Mail: talebi@physik.uni-kiel.de

**Abstract** – Ultrafast electron-photon spectroscopy in electron microscopes commonly requires ultrafast laser setups. Photoemission from an engineered electron source is used to generate pulsed electrons, interacting with a sample that is excited by the ultrafast laser pulse at a specified time delay. Thus, developing an ultrafast electron microscope demands the exploitation of extrinsic laser excitations and complex synchronization schemes.

Here, we present an inverse approach based on cathodoluminescence spectroscopy to introduce internal radiation sources in an electron microscope. Our method is based on a sequential interaction of the electron beam with an electron-driven photon source (EDPHS) and the investigated sample. An electron-driven photon source in an electron microscope generates phase-locked photons that are mutually coherent with the near-field distribution of the swift electron. Due to their different velocities, one can readily change the delay between the photons and electrons arriving at the sample by changing the distance between the EDPHS and the sample. We demonstrate the mutual coherence between the radiations from the EDPHS and the sample by performing interferometry with a combined system of an EDPHS and a $WSe_2$ flake. We assert the mutual frequency and momentum-dependent correlation of the EDPHS and sample radiation, and determine experimentally the degree of mutual coherence of up to 27%. This level of mutual coherence allows us to perform spectral interferometry with an electron microscope.

Our method has the advantage of being simple, compact and operating with continuous electron beams. It will open the door to local electron-photon correlation spectroscopy of quantum materials, single photon systems, and coherent exciton-polaritonic samples with nanometric resolution.

With the advent of ultrafast electron microscopy[1,2], visualizing the photoinduced dynamics in materials such as of magnetic vortices[3] and chemical reactions[4] has become possible at unprecedented spatial resolution. Particularly the ability to track the ultrafast dynamics of localized and propagating plasmons[5] in nano-optical systems as well as phonon polaritons in quantum materials[6] has recently boosted the application of ultrafast electron microscopy in the form of photon-induced near-field electron microscopy (PINEM)[7]. Moreover, PINEM has evolved into a unique tool for tailoring the quantum-path interferences in an extremely controllable system of single-electron wavepackets interacting with optical near fields, prepared in either classical or quantum states[8-11]. Combining real- and reciprocal-space information with elastic and inelastic processes in diffraction and electron energy-loss spectroscopy, full information about the fundamental aspects of electron-photon interactions is obtained, either in transmission or scanning electron microscopy[12-15].

In a PINEM setup, pulsed electron beams are commonly generated by virtue of the photoemission process: An ultrafast laser pulse is used to excite the apex of a sharp tip or other forms of cathodes, generating an electron pulse with a specific degree of spatial coherence, which depends on the electron source. A second laser pulse is then used to coherently induce a polarization in the sample at a certain delay with respect to the electron pulse. The stimulated interaction of the electron pulse with the laser-induced near-field excitations leads to the predominantly longitudinal modulation of the electron beams. The near-field zone of the sample hence mediates the transfer of energy and momentum from the coherent laser beam to the sample, where the strength of the photon-electron interactions is controlled by the synchronicity between the near-field excitation and the moving electron wavepacket[16-19].

Yet, a visionary application of ultrafast electron microscopy is to *coherently* control the material electronic excitations. Due to the high spatial resolution of electron-beam-based characterization techniques, electron beams could be used to coherently drive individual quantum systems to higher states[20] or to probe strong-coupling effects[21] or atomic Floquet dynamics[22] in two-level or multi-level quantum systems. Combined with mutually coherent radiation sources, quantum walks on quantized states of a quantum system, such as quantum dots, defect centers, and excitonic systems, could be coherently controlled, by initiating a set of quantum-interference paths. Coherent control methods thus require a drastic improvement of the ultrafast electron microscopy setups, to enhance the mutual coherence between the photons and electrons, such that spectral phases can be retrieved. The latter is crucial for example for retrieving quantum interference effects.

To improve the mutual coherence between photon and electron excitations in an electron microscope, we here propose and experimentally realize a proof-of-concept experiment for an inverse approach, based on intrinsic radiation emitted from the electron beam, rather than extrinsic laser radiation, to generate photons that are *phase-locked* to the near-field distribution of the swift electron. In our setup, an electron beam excites a nanostructured electron-driven photon source (EDPHS)[23-25], which generates well-collimated photon pulses (Fig. 1a and b), as shown by the angle-resolved cathodoluminescence (CL) pattern (Fig. 1c). The EDPHS consists of an array of nanopinholes in a 40 nm gold film deposited on a $Si_3N_4$ membrane created by focused ion-beam milling, with the hole radii gradually varying from 25 nm (holes in the inner rim) to 150 nm (holes in the outer rim). This allows for the generation of broadband photonic radiation (Fig. 1d)[26,27]. The radially propagating surface plasmon polaritons induced by the impacting electrons scatter off the nanopinholes and radiate into the far-field in the form of a $TM_z$-polarized Gaussian wave (see Supplementary Note 1 for a full characterization of the EDPHS radiation).

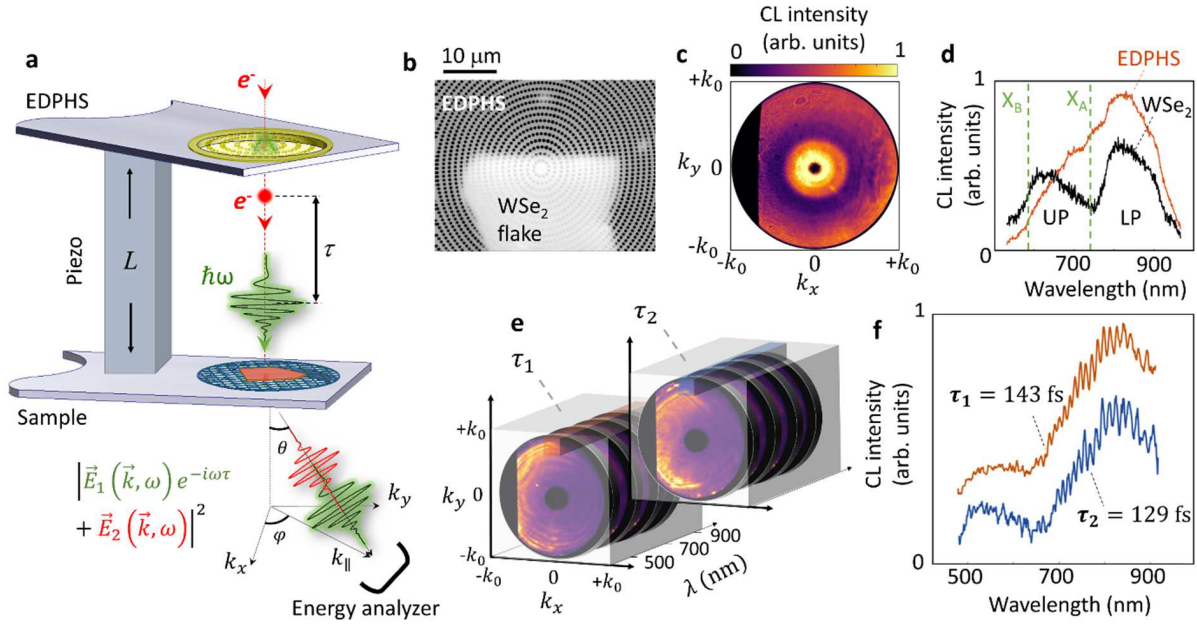

**Fig. 1. Spectral interferometry with an electron beam.** (**a**) An electron moving at a kinetic energy of 30 keV interacts with an electron-driven photon source (EDPHS) that generates photons with a collimated Gaussian spatial profile. The delay $\tau$ between photon and electron beams arriving at the sample is controlled by the distance $L$ between the sample and the EDPHS. The energy-momentum distribution of the total scattered field from the sample will be detected and analyzed to specify the mutual correlation between the EDPHS and the sample radiation, as specified in the text. (**b**) SEM image of the combination of the EDPHS and sample at the distance of $L = 2$ μm. (**c**) Angle-resolved CL pattern of the EDPHS structure. (**d**) CL spectra of the EDPHS and the WSe$_2$ flake integrated over the whole momentum space. (**e**) Schematic illustration of three-dimensional data cubes (CL intensities versus the wave numbers ($k_x$,$k_y$) and wavelength ($\lambda$)) at selected delays between the EDPHS and the sample. The shown data at the forefront are angle-resolved CL maps of the total field at the filtered wavelength of 800 nm $\pm$ 10 nm. (**f**) CL spectra of the total field at $\theta = 45° \pm 2°$ and $\varphi = 100° \pm 2°$. Both (e) and (f) maps are acquired at indicated delay values of $\tau_1 = 129$ fs ($L = 19$ μm) and $\tau_2 = 143$ fs ($L = 21$ μm).

Due to the difference between the electron group velocity ($v$) and the vacuum speed of light ($c$), the delay between the electrons and photons arriving at the sample can be precisely controlled using a piezo stage, inserted inside the sample chamber of a scanning electron microscope (SEM). Notably, our setup has 6 degrees of freedom, allowing to move both sample and EDPHs structures independently by using two nano-positioning stages (see Supplementary note 2 and Fig. S2). Specifically, the delay can be described as $\tau = L(v^{-1} - c^{-1})$, where $L$ is the distance between the EDPHS and the sample. Given an electron kinetic energy of 30 keV in our experiments ($v = 0.328\,c$), the delay can be varied in steps of 6.8 attoseconds by changing the distance $L$ in steps of 1 nm. Moreover, given the dynamic range of 6 mm for the piezo stage, the delay can be tuned within the range of 0 fs (corresponding to zero distance or touching point) to 40.8 ps.

For our proof-of-concept experiment, we use thin exfoliated WSe$_2$ flakes (80 nm thickness) placed on top of a holey carbon transmission electron microscopy grid. Fig. 1b shows the SEM image of the WSe$_2$ flake positioned blow the EDPHS, at a distance of $L = 2$ µm from the EDPHS. Belonging to the class of semiconducting transition-metal dichalcogenides, WSe$_2$ hosts energetically different A and B excitons at room temperature due to spin-orbit coupling, at energies of 1.68 eV ($\lambda = 738$ nm) and 2.05 eV ($\lambda = 605$ nm), respectively. It has previously been demonstrated that the excitons can strongly couple to the photonic modes of thin transition-metal dichalcogenide flakes[28,29], where this phenomenon is generally referred to as the self-hybridization effect[29] (Fig. S4). This coupling results in an energy splitting and opening of a bandgap, apparent in the dispersion diagram of the guided waves (Supplementary Fig. S5e), as well as the creation of lower polariton (LP) and upper polariton (UP) branches that become apparent in the CL spectrum (Fig. 1d)[30,31]. Besides optics-based characterization techniques, electron beams can be used to probe the propagation dynamics and resulting spatial coherence of exciton polaritons in thin WSe$_2$ flakes using CL spectroscopy[32]. Indeed, electron-beam spectroscopy has been applied intensively to investigate the interaction between excitons and photons or plasmons in hybrid structures[21,33,34]. Here, we develop a CL-based technique that allows us to fully determine the amplitude and phase of the aforementioned excitations. More details about the exciton polariton excitations and probing them using CL can be found in Supplementary Note S3.

Using CL, the interference between the transition radiation with the scattering of the exciton polaritons from the edges of the flakes can be used to determine the phase constant of the exciton polaritons, i.e., the change in the phase per unit length of propagation of exciton polaritons. When a moving electron approaches the surface of a flake, an image charge is induced inside the flake that together with the electron forms a time varying dipole. Its annihilation, when the electron crosses the surface, causes ultrabroadband transition radiation. The strong exciton-photon coupling and resulting energy splitting are apparent from the CL spectra of the WSe$_2$ flake (see Supplementary Note S3 for more details about the CL of the WSe$_2$ flakes). Thus, the electron beam can excite both the LP and UP branches, whereas the EDPHS radiation can excite most efficiently the LP branch (Fig. 1d). The overall detected CL signal, is the superposition of the electron-beam induced and the EDPHS-induced scattered field from the sample (Fig. 1a, bottom). The degree of mutual coherence between the elements of this superposition, is inferred from the results of an interferometry technique outlined here, and discussed step by step below.

The total momentum-resolved CL spectrum is thus rewritten as[35]

$$\Gamma_{\text{CL}}(k_\parallel,\omega) = (4\pi\hbar k_0)^{-1} \left\{ \left|\vec{E}_{\text{EDPHS}}(k_\parallel,\omega)\right|^2 + \left|\vec{E}_{\text{el}}(k_\parallel,\omega)\right|^2 + \right.$$
$$\left. \vec{E}_{\text{el}}(k_\parallel,\omega)\cdot\vec{E}^*_{\text{EDPHS}}(k_\parallel,\omega)e^{i\omega\tau} + \vec{E}^*_{\text{el}}(k_\parallel,\omega)\cdot\vec{E}_{\text{EDPHS}}(k_\parallel,\omega)e^{-i\omega\tau} \right\} \quad (1)$$

where $\hbar$ is the reduced Planck's constant, $k_0 = \omega/c$ is the free-space wave number of the light, $k_\parallel = k_0 \sin\theta = \left(k_x^2 + k_y^2\right)^{1/2}$ is the parallel wavenumber, $\vec{E}_{\text{el}}(\vec{k}_\parallel,\omega)$ and $\vec{E}_{\text{EDPHS}}(\vec{k}_\parallel,\omega)$ are the electron-induced and EDPHS-induced electric field components detected in the far field. Hence, by changing the distance between the EDPHS and the sample, the observed interference fringes in both frequency and momentum space can be controlled. The visibility of the interference fringes allows for the determination of the degree of mutual coherence between the EDPHS and electron-induced radiations. In addition, spectral interferometry is widely used to characterize the broadening of the ultrafast laser pulses. There, due to the beaming characteristic of the laser pulses, one could only measure the spectra along the longitudinal direction ($\theta = 0$)[36]. Here, despite the beaming characteristic of the EDPHS radiation, transition radiation and electron-induced radiation in general, scatter to relatively high polar angles. In contrast to collimated laser beams, the coherent electron-induced radiation pattern normally is a dipolar one[37] so that angle-resolved spectroscopy is required to fully capture the mutual coherence between EDPHS and sample radiation. Hence, the interference patterns in both momentum and energy space are characterized, and a three-dimensional energy-momentum data cube is acquired, in dependence of the delay $\tau$ between EDPHS and sample radiations. For example, by spectrally filtering the total radiation at $\lambda = 800$ nm $\pm$ 10 nm, the interference maps in the momentum space were observed and analyzed (Fig. 1e). Similarly, by filtering the angular distribution of the detected CL signal around $\theta = 45° \pm 2°$ and $\varphi = 100° \pm 2°$, corresponding to $k_\parallel = 0.707\ k_0$ and the azimuthal direction normal to the edge of the flake, the spectral interference fringes can be examined (Fig. 1f). Particularly, we notice that when the WSe$_2$ flake is excited with both the electron beam and the EDPHS radiation, the overall CL spectrum differs from an incoherent superposition of EDPHS and sample CL spectra. More importantly, the total CL angle-resolved maps and spectra vary with the distance *L* between the EDPHS and the sample (Fig. 1e and f), and delay-dependent *k*-space or spectral interferences are observed. Noteworthy, such interference patterns are only observed when the EDPH radiation as well as the electron-induced polarization inside the sample show coherent radiation properties. Therefore, polariton excitations such as plasmon polaritons in the EDPHS and exciton polaritons of the sample could be used for examining the functionality of our approach.

Since the photon generation process relies on the electron-induced surface plasmon polaritons inside the EDPHS, we anticipate that the EDPHS radiation has a high degree of mutual coherence with respect to the evanescent field accompanying the electron. Direct proof of this hypothesis is performed by using subsequent interactions of the electron beam with two similar EDPHS structures, as shown in Supplementary Note S1 and the figures therein. In order to explore the mutual coherence of the EDPHS and the sample radiation, we analyze the dependence of the angle-resolved CL patterns on *L*, by choosing a WSe$_2$ flake as the sample (Fig. 2a). The mutual correlation function between the EDPHS radiation and the sample radiation is a function of both wavelength and momentum, as stated above. First, we analyze the correlation between the EDPHS and sample radiation by filtering the overall radiation at the carrier wavelength of the EDPHS radiation (i.e., $\lambda = 800$ nm) and analyzing the angle-resolved radiation pattern. When only the electron beam excites the sample, specific interference fringes in the angle-resolved CL pattern are observed, due to the interference between the transition radiation and exciton polaritons (Fig. 2b) (see Supplementary Note S3). A drastic alteration of the interference fringes is observed when the EDPHS radiation and the electron beam simultaneously excite the sample. The momentum-distance interference fringes, are observed within the distance range of $L = 22$ μm to $L = 40$ μm, (Fig. 2c, Fig. 2d, region $R_1$), and these interference fringes are altered by including the EDPHS radiation, which is determined by the temporal coherence of the generated EDPHS radiation and decoherence phenomena, due to the interaction of the superimposed EDPHS and the sample radiation with the environment. This latter effect is precisely the reason why the visibility of the interference fringes is diminished by further increasing the distance *L* above 40 μm. Performing the measurements in finer steps, we are able to resolve the interference fringes versus the transverse angular momentum and distance *L*, *demonstrating the high degree of mutual coherence between the EDPHS and the sample radiation*. The interference fringes – within the fully coherent range - can be simulated using classical electromagnetism considering a realistic system of the EDPHS radiation and the sample (Fig. 3) (see Supplementary Notes S3 and S4 for details of the simulation method). In the simulation, a rectangular WSe$_2$ flake is considered that is sequentially excited by a swift electron at the kinetic energy of 30 keV and the EDPHS radiation (the previously performed simulation results, which includes the interaction of the electron beam with our EDPHS structure, is used (see Fig. S1)). The EDPHS-induced and electron-induced polarizations are superimposed at the corresponding delays (Fig. 3a and b) and the far-field radiation is obtained by projecting the field distributions from the near-field to the far field, using free-space Green's functions. The delay between the electron-induced and EDPHS-induced polarization affects the total diffraction angle of the field. Instead of the superposition of two waves with the same propagation direction (as for the combination of two identical EDPHS structure (Supplementary Fig. S5)), a directional beam and dipolar field profile are superimposed. The latter radiation is caused by the transition radiation mechanism as an example. Thus, the interference patterns are highly momentum-dependent. The agreement between the simulation and experimental results suggests that the superposition of the EDPHS and electron-induced scattered field from the sample underpin the experimentally observed interference patterns.

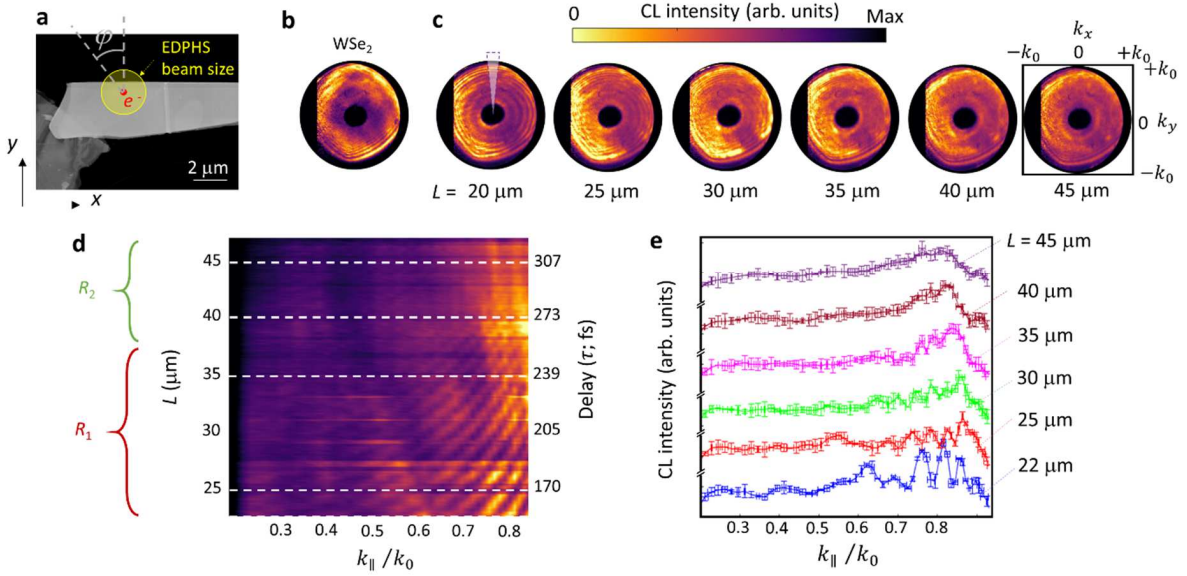

**Fig. 2. Angle-resolved cathodoluminescence maps at the filtered wavelength of $\lambda = 800$ nm versus the delay $\tau$ between the EDPHS and the sample.** (a) SEM image of the WSe$_2$ flake. The EDPHS beam size and the electron impact position are both depicted on the image. Angle-resolved CL maps of (b) the WSe$_2$ flake and (c) the combination of the WSe$_2$ flake and EDPHS at indicated distances between them ($k_x = k_0 \sin\theta\cos\varphi$ and $k_y = k_0\sin\theta\sin\varphi$, where $\theta$ and $\varphi$ are the polar and azimuthal angles with respect to the sample plane). Electrons traverse the flake at a distance of 800 nm from the edge of the flake. (d) Measured $L-k$ CL map acquired at the azimuthal angle range marked in (c) with a purple triangle. Here, $k_\parallel^2 = k_x^2 + k_y^2$. The mutual spatial coherence between the EDPHS and CL radiation is demonstrated by the visibility of interference fringes, in the region $R_1$. $R_2$ denotes the region where the visibility of the interference fringes vanishes. All angle-resolved CL maps are taken at a wavelength of 800 nm. Dashed lines in the left panel indicate the distances at which the full angle-resolved maps were acquired as shown in (c) and at which the CL intensity-$k$ line plots are shown in (d). For complete visualization of the interference maps, see Supplementary Movie 1.

To better understand this effect, we propose a model to reconstruct the interference patterns using geometrical optics (Fig. 3d; lower right panel). For this, we consider possible beam paths that contribute to the far-field patterns as (i) the EDPHS radiation that is directly transmitted through the film, (ii) the EDPHS radiation that is scattered off the edge of the flake, (iii) the transition radiation, and (iv) the excitation of the exciton polaritons and their scattering from the edges. First, we notice that EDPHS excitation cannot directly excite the exciton polaritons, due to the momentum mismatch between the exciton polariton dispersion and that of the free-space light. The scattering of the EDPHS radiation from the edges results in the excitation of exciton polaritons and forms a standing wave pattern inside the film that also contributes to the scattered light from the edges (see Fig. 3a for a visualization of the standing-wave pattern and its scattering from the edges). In addition, the electron beam directly excites the exciton polaritons as well and also causes transition radiation (see Supplementary Note S3). The interferences between these four beam paths form the interference pattern depicted in Fig. 3d in the lower left panel, matching the simulated and experimental results. Minor disagreements are due to the fact that in the model, only scattering from two edges are included, whereas the experimental and simulation results include more scattering edges.

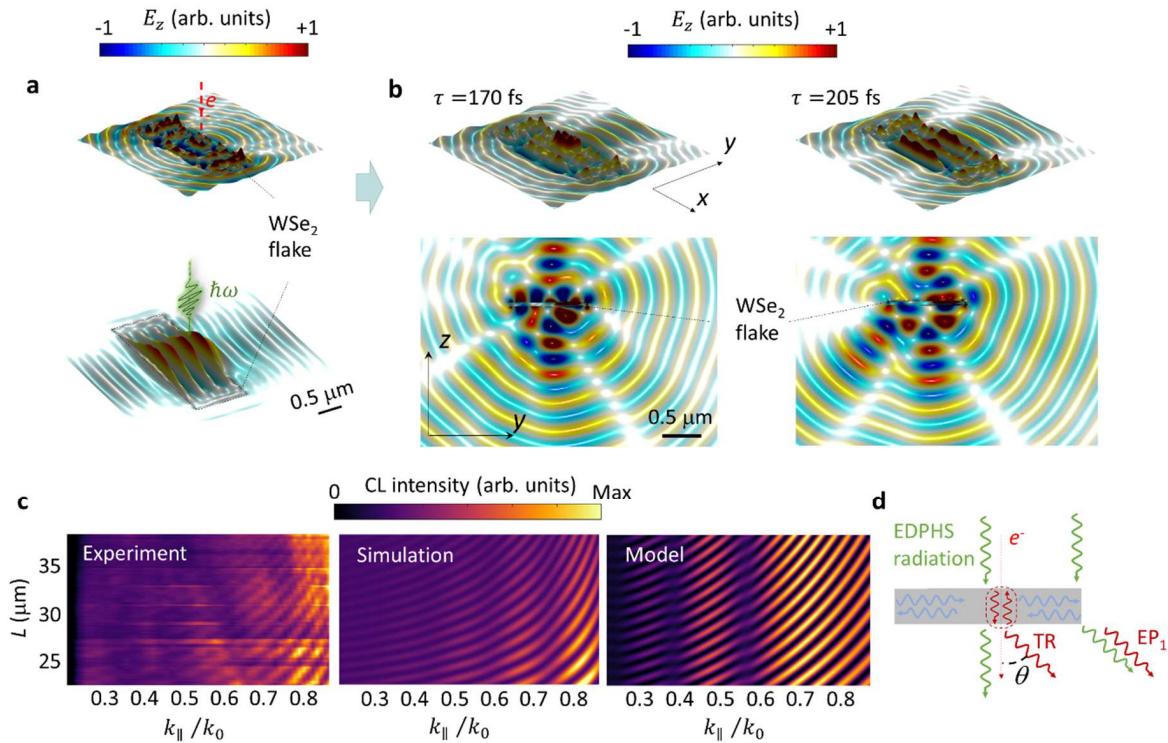

**Fig. 3. Modeling the experimentally observed *k*-space interference fringes.** (a) Simulated near-field distributions induced by a moving electron at the kinetic energy of 30 keV (top), and the EDPHS radiation (bottom), and (b) their superposition at two depicted delay times between the EDPHS radiation and the coming electron beam, at the *x-y* plane located 5 nm above the sample (top row), and the *y-z* cross section, cutting parallel to the electron beam trajectory at the electron beam impact position (bottom row). (c) (Left) Experimental results compared to the (middle) simulated CL intensity versus in the momentum – distance space (distance between the EDPHS and the sample, *L*, and the transverse momentum parallel to the shorter symmetry axis of the flake), compared to the analytical results (right) obtained using the (d) proposed model that considers the interference between a direct transmission of the EDPHS radiation through the film and its scattering from the edge, the transition radiation (TR), and the scattering of the excited exciton polaritons (EPs) from the edges. Contributions of the EDPHS and the electron-induced radiation are shown by green and red wavy arrows, respectively.

The model thus reproduces the measured pattern, further confirming the high degree of mutual coherence between the EDPHS and the sample radiation. Comparing the experimental results with the results of classical electromagnetic simulations, a degree of coherence of 27% is inferred. Increasing the spatial distance to $L > 40$ μm, leads to the deterioration of the visibility of the interference fringes (Fig. 2d, Region $R_3$). This behavior is a peculiar example of the decoherence phenomena, where the interaction of the individual components of the radiation field with the environment suppress the coherence of the EDPHS and the electron-induced polarization in the sample (see Supplementary Note S4 and Supplementary Fig. S6 for more information).

The high degree of coherence between the EDPHS and the sample radiation, within the aforementioned distance range, is also spectrally analyzed and used for spectral interferometry, as we demonstrate in the following. The angle(momentum)-resolved CL spectra of the combined EDPHS and sample radiation shows as well a clear interference map, at higher momentum ranges between $k_\parallel = 0.7\ k_0$ and $k_\parallel = 0.8\ k_0$ (Fig. 4a). The LP branch is prominently excited, as the EDPHS radiation peaks at 800 nm. Moreover, by changing the delay (distance between the EDPHS and sample), both the modulation frequency as well as the visibility of the interferences fringes are altered (Fig. 4b). The dependency of the CL signal on $k_\parallel$, using the technique compared here, i.e., filtering the signal along the azimuthal degree of freedom using a mechanical slit, is compared to the signal obtained beforehand, where the angle-resolved patterns were obtained at the filtered wavelength of $\lambda = 800$ nm (Fig. 4c). Obviously, good agreement is obtained. These spectral interference fringes are thus used to retrieve the spectral phase in the following.

To proceed, we first rewrite eq. (1) into three components as[38]

$$\Gamma_{\text{CL}}(\omega) = |I_0(\omega)| \left\{ 1 + |\sigma(\omega)|^2 + \sigma(\omega) e^{i\omega\tau} + \sigma^*(\omega) e^{-i\omega\tau} \right\} \tag{2}$$

at a fixed $k_\parallel$ value, where $\sigma(\omega) = E_{z,\text{el}}(\omega) / E_{z,\text{EDPHS}}(\omega)$ is the ratio of the EDPHS and the sample $z$-components of the electric field radiated to the far field, and $|I_0(\omega)| = (4\pi\hbar k_0)^{-1} |\vec{E}_{\text{EDPHS}}(k_\parallel, \omega)|^2$. The first term, i.e., $\Gamma_0(\omega) = |I_0(\omega)| \left\{ 1 + |\sigma(\omega)|^2 \right\}$, does not have any dependence on the delay $\tau$, whereas the last two terms, defined as $\Gamma_+(\omega) = |I_0(\omega)| \sigma(\omega) e^{+i\omega\tau}$ and $\Gamma_-(\omega) = |I_0(\omega)| \sigma^*(\omega) e^{-i\omega\tau}$, clearly depend on the delay. Thus, taking the Fourier transform of the overall CL spectrum, one can transfer all the components into the time domain, with $\tilde{\Gamma}_0(t) = \Im\{\Gamma_0(\omega)\}$ and $\tilde{\Gamma}_\pm(t) = \Im\{\Gamma_\pm(\omega)\}$ centering at $t = 0$, and $t = \pm\tau$, respectively. Now we perform this for the CL signal at the distances of $L = 20$ μm, $L = 22$ μm, and $L = 30$ μm, all at $k_\parallel = 0.77\, k_0$, where the time-dependent CL signal is obtained. Three dominant peaks are observed as expected, namely, the DC term at $t = 0$, and the AC terms at $t = \pm 136$ fs for $L = 20$ μm, $t = \pm 150$ fs for $L = 22$ μm, and $L = \pm 204$ fs for $L = 30$ μm, respectively. The occurrence of the DC term is clearly due to the delay-independent CL intensities corresponding to the individual EDPHS and sample CL signals ($\tilde{\Gamma}_0(t)$; the first two terms on the RHS of equation (1)). In addition, the AC terms that occur exactly at $\tau = L(v_{el}^{-1} - c^{-1})$, are due to the last terms on the RHS of eq. (1). Assuming that the CL intensities from the EDPHS and the sample are at the same level of strength, the degree of mutual coherence is obtained as $CL^{AC}/CL^{DC}$, which corresponds to exactly 27%, as inferred from the comparison of the simulated and measured results before. The ratio of the CL intensities of the EDPHS and the sample is experimentally confirmed by taking the CL spectra of individual components under the same experimental conditions (see Fig. 1d). The broadening of the DC signal is also acquired by taking the bandwidth of the DC peak, which corresponds to 5.2 fs FWHM. Thus, the temporal broadening of the EDPHS and sample radiation are both approximately 5.2 fs. Taking the inverse Fourier transform of only the AC signal and by filtering the time-dependent signal around the AC peak, the spectral amplitude and phase are both retrieved (Fig. 4c). For doing this, we note that

$$|\Gamma_0(\omega)|/|\Gamma_\pm(\omega)| = \left(1+|\sigma(\omega)|^2\right)/|\sigma(\omega)| \tag{3}$$

allowing us to compute the relative electric-field amplitude as

$$|\sigma(\omega)| = \frac{|\Gamma_0(\omega)|}{2|\Gamma_+(\omega)|} - \sqrt{\left(\frac{|\Gamma_0(\omega)|}{2|\Gamma_+(\omega)|}\right)^2 - 1}. \tag{4}$$

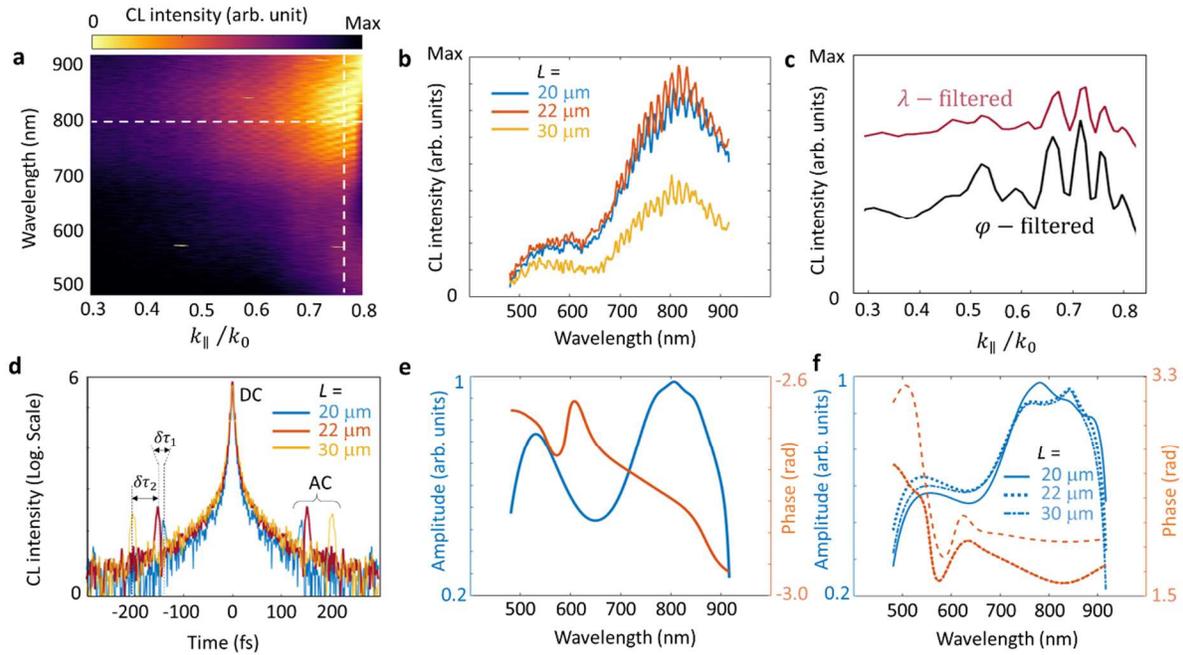

**Fig. 4. Spectral interference fringes.** (a) Momentum-resolved CL spectra at the distance of $L = 22$ μm (delay $\tau = 150$ fs), for an electron traversing the WSe$_2$ flake at a distance of 2 μm from the edge. (b) CL intensity acquired at the wavenumber of $k_\parallel = 0.77\, k_0$ at depicted distances. (c) CL intensity versus the lateral momentum, where the results obtained by filtering along the azimuthal direction via a mechanical slit and then selecting the spectral content at $\lambda = 800$ nm, are compared to the results obtained by spectrally filtering the radiation at $\lambda = 800$ nm, and then selecting the azimuthal range $\varphi = 90° \pm 2$. (d) Fourier-transformed CL intensity at depicted L values, with three peaks at $t = 0$ (DC term) and $t = \pm\tau$, at depicted distances between the sample and the EDPHS. $\delta\tau_1 = 14$ fs and $\delta\tau_2 = 55$ fs correspond to $\delta L = 2$ μm and $\delta L = 8$ μm, respectively. Retrieved relative electric-field amplitude and phase with respect to the EDPHS for (e) $k_\parallel = 0.77\, k_0$, using the CL signal acquired for $L = 22$ μm, and (f) $k_\parallel = 0.7\, k_0$, using the CL signal acquired for different L values.

Moreover, the relative phase is obtained by simply retrieving the phase of $\Gamma_+(\omega)$, as $|I_0(\omega)|$ is a real-valued quantity. Moreover, since the CL spectrum of only the EDPHS radiation is easily obtained at the first stage, the electric field amplitude of only the sample radiation can be retrieved. However, only the differential CL phase between the sample and EDPHS radiation can be acquired with this technique, since no information about the phase of the EDPHS radiation is at hand. The proposed phase-retrieval algorithm should not depend on the delay between the reference and the signal, as far as the AC and DC spectral components are completely distinguishable. This fact can be used as a benchmark for obtaining the accuracy of the proposed technique. Retrieving the intensity and phase at different $L$-values, the fluctuations in the obtained results are negligible for $L = 20$ μm and $L = 22$ μm. However, a maximum difference of the phase value in the order of 20% is obtained, when comparing the values for $L = 20$ μm and $L = 30$ μm, which provides an estimate for the accuracy of our spectral interferometry technique.

Thus, the proposed algorithm can be used to retrieve the amplitude ($|\sigma(\omega)|$) and the phase ($\alpha(\omega)$) of the momentum-energy maps (Fig. 5). The accuracy of the acquired maps depend on the visibility of the interference fringes, thus are reliable for $0.5k_0 \leq k_\parallel \leq 0.8k_0$. The retrieved amplitude (Fig. 5a) shows a smooth shift of the LP and UP polariton branches toward shorter wavelengths upon increasing the transverse momentum. This behavior is expected from the dispersion of exciton polaritons (strong exciton-photon interactions), for both LP and UP branches. In contrast, the retrieved phase shows fluctuating behavior versus transverse momentum and a less obvious fluctuation versus wavelength. Considering the lowest-order scattered rays, (see Supplementary Note 4), the relative phase can be described as $\alpha_{n=0}(\omega) = k_\parallel l_1 - \beta(\omega) l_1 - \varphi_{T_2}(\omega)$, which is a smooth function of $k_\parallel$. Thus, we notice that the fast fluctuation behavior of the relative phase is due to the inclusion of higher order scattering terms.

The correlative photon-electron spectroscopy based on EDPHS thus allows for *phase-stable* spectral interferometry by improving the mutual coherence between the arriving photons and electrons at the sample. Moreover, the compactness of the design allows to minimize decoherence in the photon-electron interaction because the photon-generation process happens at a distance of only a few micrometers above the sample. The scheme thus maximizes the mutual coherence between the electrons and photons. Intriguingly, advanced nanofabrication techniques could be used to design EDPHS structures with tailored photon emission properties. Generating vortex light or even temporally shaped optical pulses is possible by the control of multiscattering events and engineering of defect centers in both the lateral and longitudinal directions. The approach thus opens new directions in understanding the momentum-spectral correlations in polaritonic materials and correlated electron systems such as transition-metal dichalcogenides. Merging this method with advanced holographic techniques[39] allows for unravelling a variety of information about the charge and energy transfer dynamics at ultimately attosecond time resolution. Moreover, this setup has the potential for exploring fundamental aspects of electron-photon interactions and addressing key questions such as entanglement between generated photons from different scattering events, which could be addressed by combining this with an electron-beam analyzer and spectrometer[40] in an SEM[15].

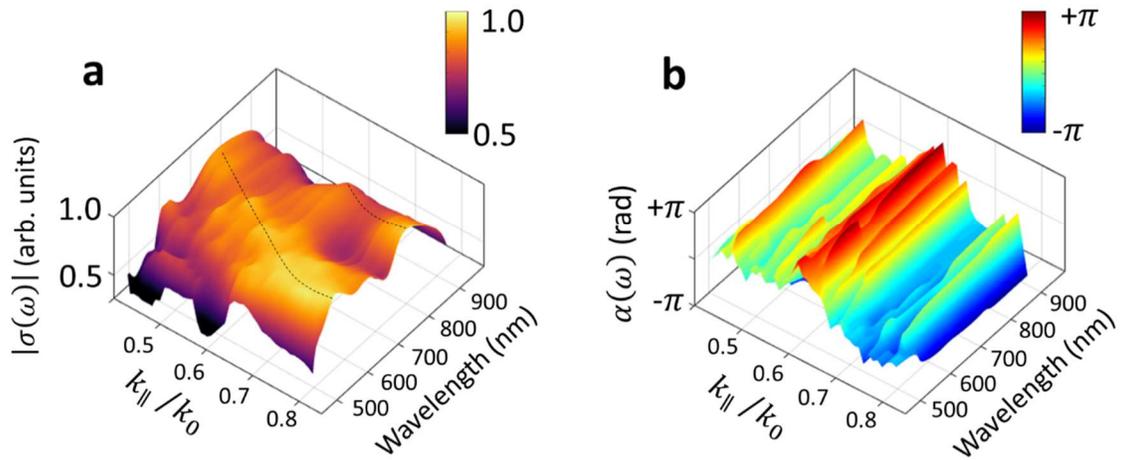

**Fig. 5. Momentum-wavelength map of the relative EDPHS- and electron-beam-excitations amplitude and phase.** Retrieved (a) amplitude, and (b) phase.

**Acknowledgements:**

N.T. acknowledges fruitful discussions with Christoph Lienau and Claus Ropers. This project has received funding from the European Research Council (ERC) under the European Union's Horizon 2020 research and innovation program, Grant Agreement No. 802130 (Kiel, NanoBeam) and Grant Agreement No. 101017720 (EBEAM). M.H. and H.G. thank DFG, BMBF, and ERC Grant (ComplexPlas) for funding.

**Supplementary Material:**

# Phase-locked photon-electron interaction without a laser

Masoud Taleb[1,2], Mario Hentschel[3], Kai Rossnagel[1,2,4], Harald Giessen[2], and Nahid Talebi[1,2*]

[1]*Institute of Experimental and Applied Physics, Kiel University, 24098 Kiel, Germany*

[2]*Kiel Nano, Surface and Interface Science KiNSIS, Kiel University, 24118 Kiel, Germany*

[3]*4th Physics Institute and Research Center SCoPE, University of Stuttgart, 70569 Stuttgart, Germany*

[4]*Ruprecht Haensel Laboratory, Deutsches Elektronen-Synchrotron DESY, 22607 Hamburg, Germany*

E-Mail: talebi@physik.uni-kiel.de

Content:

1. Characterizing the EDPHS radiation
2. Phase-locked photon-electron spectroscopy setup
3. Exciton polaritons in $WSe_2$ thin films
4. Mutual coherence and simulated interference map

1. **Characterizing the EDPHS radiation**

A full characterization of the EDPHS radiation generally requires pulse-characterization techniques. Here, we use a simple interferometric technique that is able to demonstrate the degree of coherence of the radiation. Before we get to the experimental results, we first show the simulation results to clarify the collimation and chirping of the radiation pattern of the EDPHS. A time-dependent finite-difference method, which is fully explained elsewhere, is used[41]. The discretization unit cell is 2 nm, and for the calculation a supercomputer at the Kiel University Computing Center is used. 60000 temporal iterations were required to cover the time period of 230 fs, thus allowing convergence. The simulation time was 108 hours.

In contrast to the EDPHS structure reported in ref.[26], here the radiation from the EDPHS is not focused, but it is collimated, as obvious from the angle-resolved pattern shown in Fig. 1 of the main text (Fig. S1a). The collimated EDPHS radiation helps to maintain the same EDPHS beam profile on the sample via changing the distance between the EDPHS and the sample, allowing to systematically investigate the role of decoherence without altering the beam divergence and profile, as shown in section 4 of the supplementary notes. Nevertheless, we have used the same design principle reported before[26], which is based on distributing a two-dimensional array of nanopinholes in a gold thin film (Fig. S1a). The distribution of the holes is determined via an inverse holographic approach[23]. This structure obviously hosts a large number of optical modes that are also energetically distributed in a closed-pack manner, resulting in an inhomogeneous broadening of the near-field spectra; consequently, the far-field radiation from the structure is chirped (Fig. S1b). Windowed Fourier transformation of the time-dependent field demonstrates the down-chirp of the optical radiation (Fig. S1c), where the wavelength of the signal increases with time. As a result, the total spectrum (black solid line) is a broadband excitation with a bandwidth of approximately 200 nm centered at 820 nm, in close agreement with the measured CL spectrum (Fig. 2d). The spatial profile of the EDPHS radiation also captures the collimation of the EDPHS radiation over a broad wavelength range from 700 nm to 1000 nm (Fig. S1d), where at shorter wavelength, the excitation of side lobes is also apparent, causing a diffused radiation pattern at larger polar angular ranges in the angle-resolved CL map.

In order to quantify the coherence length of the EDPHS radiation, we investigate the interaction of the electron beam with two identical EDPHS structures positioned along the path of the electron beam at a distance $L$ (Fig. S2a). The far-field radiation from the whole structure is the superposition of the radiations from each individual EDPHS structure, in addition to the scattered field caused by the interaction of the first EDPHS radiation with the second one (Fig. S2b). The acquired momentum-resolved CL map at the filtered wavelength of 800 nm (corresponding to the carrier wavelength of the EDPHS radiation) shows a prominent interference pattern versus the distance between the EDPHS structures, in full agreement with the simulation results (Fig. S2c and d). The agreement between the experimental and simulated results is encoded more prominently in the line profile captured by integrating the CL intensity over the whole momentum axis (Fig. S2e). The distance between the maxima of the interference pattern does not depend on the acquired transverse momentum, showing that the main contribution is caused by the interference between the optical beams that propagate along the same direction, i.e., the collimated beam paths. This effect is in contrast to the case where the WSe$_2$ flake is used as the sample. For the latter case, as will be discussed later, the interference between waves with dipolar wave fronts such as transition radiations and the scattering of the exciton polaritons from

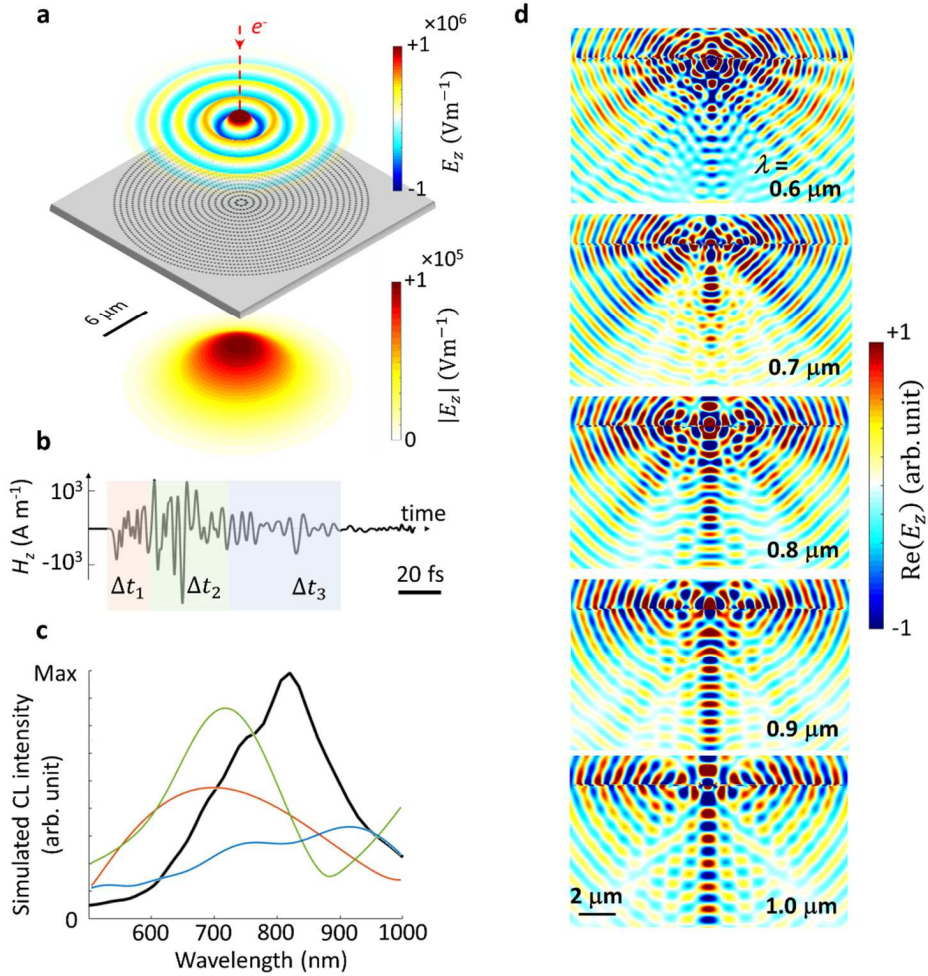

**Fig. S1. Numerical simulations showing the collimation and chirping of the EDPHS radiation.** (a) EDPHS design based on transforming a radially propagating surface plasmon polariton induced by the electron beam, into a collimated Gaussian beam, via its interaction with a distribution of nanopinholes. (b) Simulated temporal profile of the radiation at the far-field zone, demonstrating a chirp excitation, with its full spectrum (black solid lines) and spectra acquired from the colored window regions shown in (c). (d) Spatial profile of the amplitude of the z-component of the electric field at depicted wavelengths and at a given time, where the collimated radiation is obvious.

the edges, with directional waves, cause an interference pattern that is significantly $k_\parallel$ – dependent. The maximum intensity of the interference pattern is only observed after the distance of 22 µm (delay of $\tau = 148$ fs) between the EDPHS structures, which corresponds to the required time for the induced polarization in the EDPHS to substantially contribute to the radiation continuum and moreover overlap with the electron induced radiation from the second EDPHS. Since electrons are propagating at the speed of 0.328 $c$, the retardation time for the EDPHS radiation is obtained as $\delta\tau = 22\,\mu\text{m}\left(v_{\text{el}}^{-1} - c^{-1}\right) = 150\,\text{fs}$. This time onset of the observation of the interference phenomenon directly matches with the experiment

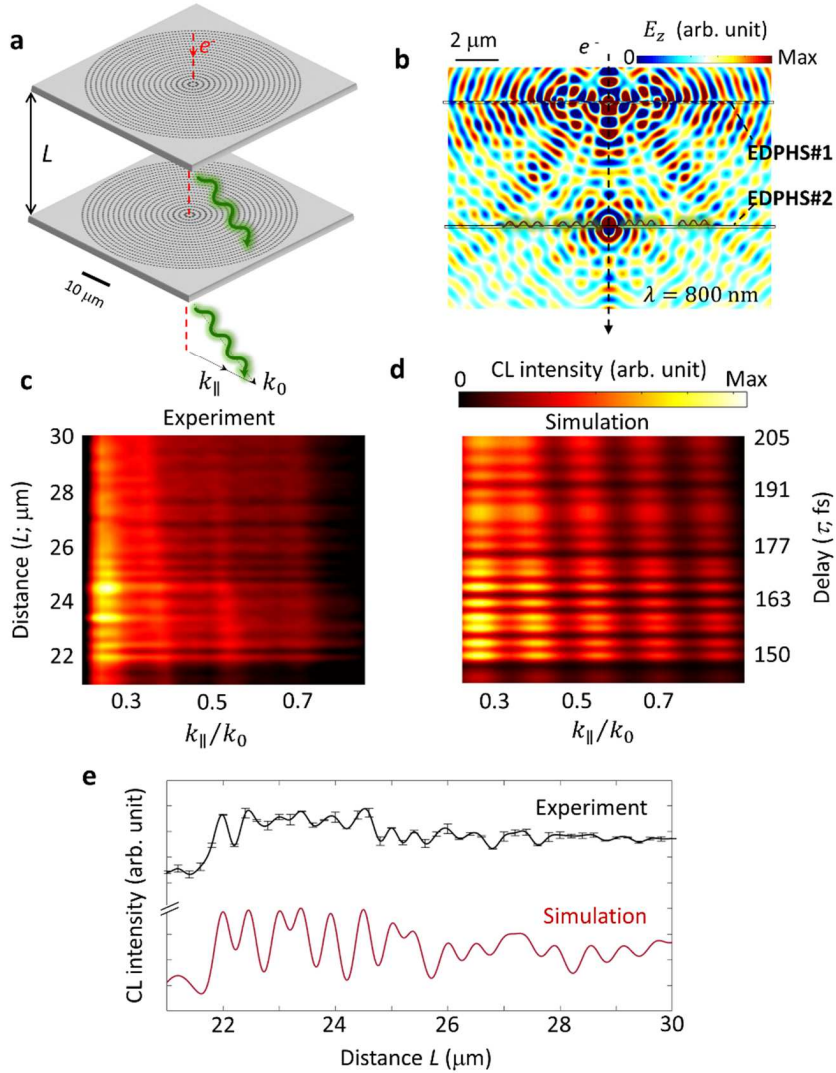

**Fig. S2. Characterizing the coherence length of the EDPHS radiation.** (a) Two similar EDPHS structures are positioned at distance *L* with respect to each other along the trajectory of the electron beam, where the electron beam interacts with both structures at a delay $L/v_{\text{el}}$. Depending on the delay, the EDPHS radiations interfere constructively or destructively in the far-field. (b) The spatial profile of the z-component of the electric field for the double-EDPHS structure, at the wavelength of 800 nm. (c) Measured and (d) simulated CL intensity versus the transverse momentum and the distance *L*. (e) Line profiles of the measured and simulated integrated CL intensities versus the distance.

that involves the WSe$_2$ flake as the second interaction point, and is linked to the retarded time required for the EDPHS structure to contribute to the radiation continuum.
15

2. **Phase-locked photon-electron spectroscopy setup**

Our phase-locked photon-electron spectroscopy setup is based on the integration of a nanopositioner system with a cathodoluminescence detection unit. We have designed a vacuum-compatible compact piezo stage in collaboration with the SmarAct Company[1], with three axial degrees of freedom, 1 nm accuracy in the step size, and a dynamic range of 12 mm (Fig. S3a and b). The electron-driven photon source (EDPHS) is held by a solid arm by the nanopositioner and its lateral and vertical positions with respect to the sample are precisely controlled. The position of the sample is controlled by the SEM sample stage (Fig. S3a and b). Our system is equipped with two parabolic mirrors that collect the photons emitted by the combined SEM/EDPHS system (Fig. 3c). The mirror positioned on top of the stage has a larger photon collection efficiency. The mirror positioned below the sample – used in the present measurements – has a larger hole in the middle, to avoid damage upon electron-beam irradiation.

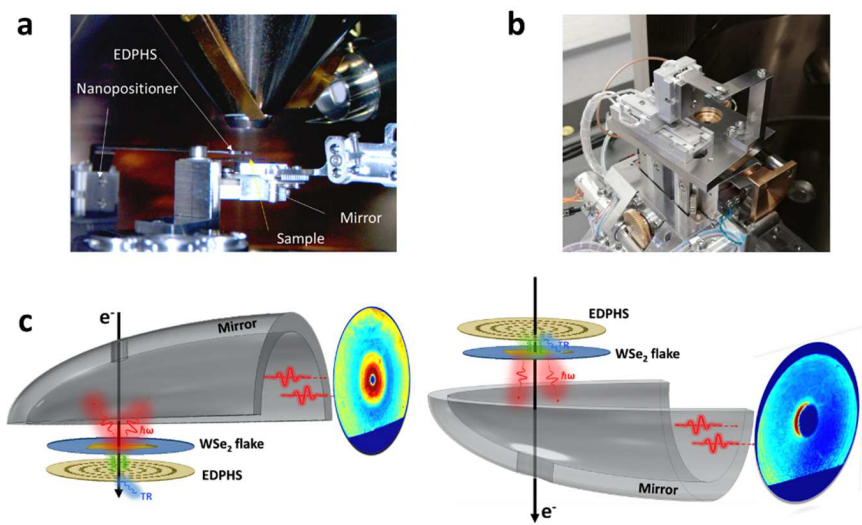

**Fig. S3. Correlative photon-electron spectroscopy setup.** (a) Schematic of the setup positioned in the chamber of the electron microscope, including a nanopositioner piezo stage for aligning the EDPHS position with respect to the electron beam and the sample. (b) Nanopositioner positioned on the sample stage of the electron microscope. (c) Collector mirrors located above (left) and below (right) the sample, and the angle-resolved CL pattern of the EDPHS structure as captured by both mirrors.

---

[1] https://www.smaract.com/index-en

### 3. Exciton polaritons in WSe$_2$ thin films

High-quality crystals of 2H-WSe$_2$ were grown by the standard chemical vapor transport method. Thin nanosheets were prepared by applying liquid phase exfoliation (LPE) from the bulk WSe$_2$ in isopropanol (Merck, ≥ 99.8%). Exfoliation was performed with an ultrasonicator (320 W, Bandelin Sonorex, RK100H) equipped with a timer and heat controller to avoid solvent evaporation. The resulting suspension was drop casted on a holey carbon mesh grid for further characterization.

Exciton-photon interactions are categorized into weak- and strong-coupling regimes, where the former happens for the scattering of light from excitons in free space, and the latter could be obtained when the excitons, normally considered as two-level systems, are positioned inside a Fabry-Perot cavity. Due to the increased interaction time, which can be estimated by the finesse of the cavity, the effective absorption cross section of the excitons is increased, and the interaction leads to new hybridized exciton-cavity states. To reach the strong-coupling regime with a number of $N_a$ emitters inside a cavity, the criterion to be met is $N_a C = N_a g^2 / 2\kappa\gamma \gg 1$, where $g = \sqrt{\mu^2 \omega / 2\varepsilon_0 \hbar V}$ is the exciton-cavity Rabi frequency, $\kappa = \pi c / 2LF$, and $\gamma = \mu^2 \omega^3 / 6\pi\varepsilon_0 \hbar c^3$ is the spontaneous decay rate of the excitons[42]. The parameter $L$, the length of the cavity (here, the thickness of the film) drops out, as it appears also in the parameter $V = L \cdot A^2$, the volume of the cavity. $F$ is the finesse of the cavity, which can be calculated by the reflection coefficients at the WSe$_2$/air interface (Fig. S4a and b). Obviously, due to the large refractive index of WSe$_2$, the criterion for the total internal reflection is easily met. Nevertheless, the material loss directly affects the performance of the cavity in general and hence the finesse is also affected by the loss inside the cavity (Fig. S4b). Considering 100 emitters within an area of $A = 1\mu m \times 1\mu m$ of the cavity, the criterion mentioned above could be met for $\lambda > 600$ nm. Moreover, the coupling efficiency increases with increasing the wavelength, providing the reason for the stronger interactions between A excitons and photons, compared to B excitons.

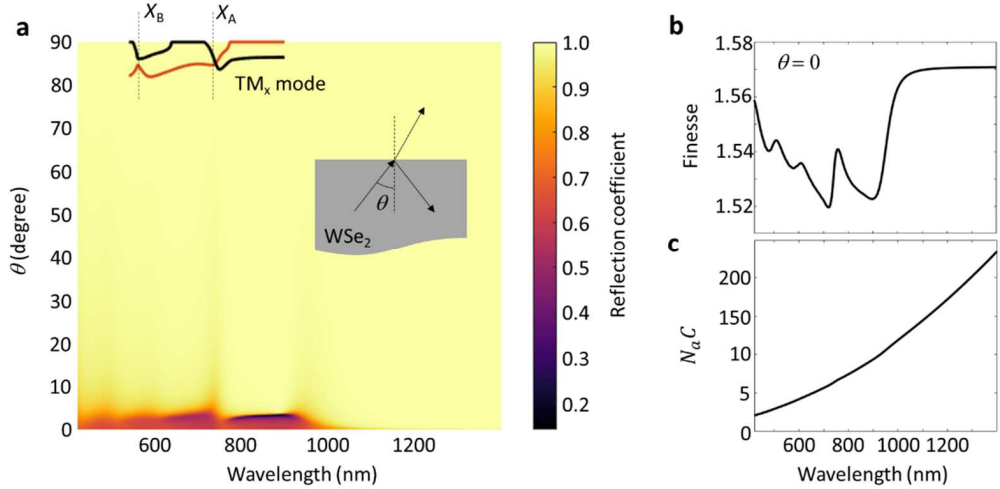

**Fig. S4. Strong-coupling criterion for exciton-photon interactions in WSe$_2$ thin films.** (a) Reflection coefficient for an interface between WSe$_2$ and air. Solid lines show the angle-wavelength dependence associated with the TM$_x$ mode of a WSe$_2$ thin film with the thickness of $d$ = 80 nm. (b) The finesse parameter associated with a symmetric WSe$_2$ film, holding Fabry-Perot like resonances at $\theta = 0$. (c) Cooperativity parameter for a number of 100 excitonic emitters inside a volume of $d \times (1\ \mu m)^2$ of the optical cavity (thin film).

Due to strong exciton-photon interactions in the WSe$_2$ flakes, lower-polariton (LP) and upper-polariton (UP) branches are formed, and the corresponding propagation mechanisms in thin films can be probed by CL spectroscopy[32] (Fig. S5a, b and c). Particularly comparing the absorption spectrum (Fig. S5d) with the CL spectrum (Fig. S5c), the strong interaction between light and the cavity modes and its signature on the CL spectra become apparent. Moreover, the interaction of the propagating polaritons with the edges of the flakes is obviously tracked by scanning the sample with the electron beam and acquiring the CL spectra at each point. Strong interactions between photons and excitons can be probed in the time domain by observing Rabi oscillations. Spectral signatures of Rabi oscillations include an energy splitting and level repulsion in the reciprocal domain that becomes apparent by the energy splitting observed in the CL spectra (Fig. S5c). The amount of the energy splitting between LP and UP branches determines the exciton-photon coupling strength that varies as a function of the distance from the edge of the flake. The dispersion of the exciton polaritons in such a thin film is analytically calculated (Fig. S5e) and perfectly demonstrates the level repulsion caused by the coupling between the photonic mode of the thin film (black dashed line) and the excitons (solid green lines). The propagating exciton polaritons are partially reflected from the edges of the flake and partially scattered to the far-field (Fig. S5f and g). Particularly, the interferences observed in the far-field angle-resolved map can be understood by the interference between the transition radiation and the scattering of the exciton polaritons from the edges of the flakes.

Transition radiation (TR) is the radiation caused by the time-varying dipole, formed by the interaction between the swift electron and its image inside the material, when the electron approaches the surface of the material. The annihilation of this dipole when the electron reaches the surface of the material creates ultrafast radiation that covers the spectral range from a few meV to several eVs[43], having a dipolar-like radiation pattern[44]. Transition radiation has been recently used to retrieve the spatial phases attributed to plasmonic excitations[45]. Within the WSe$_2$ flakes, the excited transition radiation can interfere with the scattered exciton polaritons from the edges of the flakes. Since the electron-induced exciton polaritons sustain a radial wave front, a simple model can be constructed to retrieve the far-field interference pattern, by assuming that the superposition of the scattered waves from the edges of the flake and the transition radiation can be formulated as

$$\psi(\vec{r},\theta) = \frac{e^{-ik_0 r}}{4\pi r} \sin\theta \left(1 + |T_2| e^{ik_0 \sin\theta l - i(\beta - i\alpha)l + i\varphi_{T_2}}\right). \tag{S1}$$

Therefore, the maxima of the far-field interference pattern are recast as

$$\beta(\omega)l = k_0 \sin\theta l + 2n\pi + \varphi_{T_2} \tag{S2}$$

where $\beta(\omega) - i\alpha(\omega)$ is the frequency-dependent complex-valued propagation constant of the exciton polaritons (Fig. S2e). Its real part, i.e., the phase constant $\beta(\omega)$, is the difference in phase per meter of the propagation of the polaritons in the thin film. $\alpha(\omega)$, i.e., the attenuation constant, models the attenuation of the polaritons per meter of their propagation, due to the materials dissipation or radiation loss. $|T_2|e^{i\varphi_{T_2}}$ is the complex-valued transmission coefficient of the z–component of the electric field (this term will be discussed in Supplementary Note S4), $l = p/\cos\varphi$ is the propagation length of the exciton polariton (Fig. S5f and g), with $p$ being the electron impact parameter with respect to the edge, $\varphi$ the azimuthal angle, and $\theta$ the polar angle. Considering two edges, the angular profile in the far-field zone attributed to the maxima of the constructive interference events are obtained and indicated by the contours shown in Fig. S5h, where $p = 1200\,\text{nm}$ has been considered. The position of these contours matches well with the experimentally observed interference patterns in the angle-resolved CL maps, where only up to *n* = 3 interference orders are observed (Fig. S5i and j).

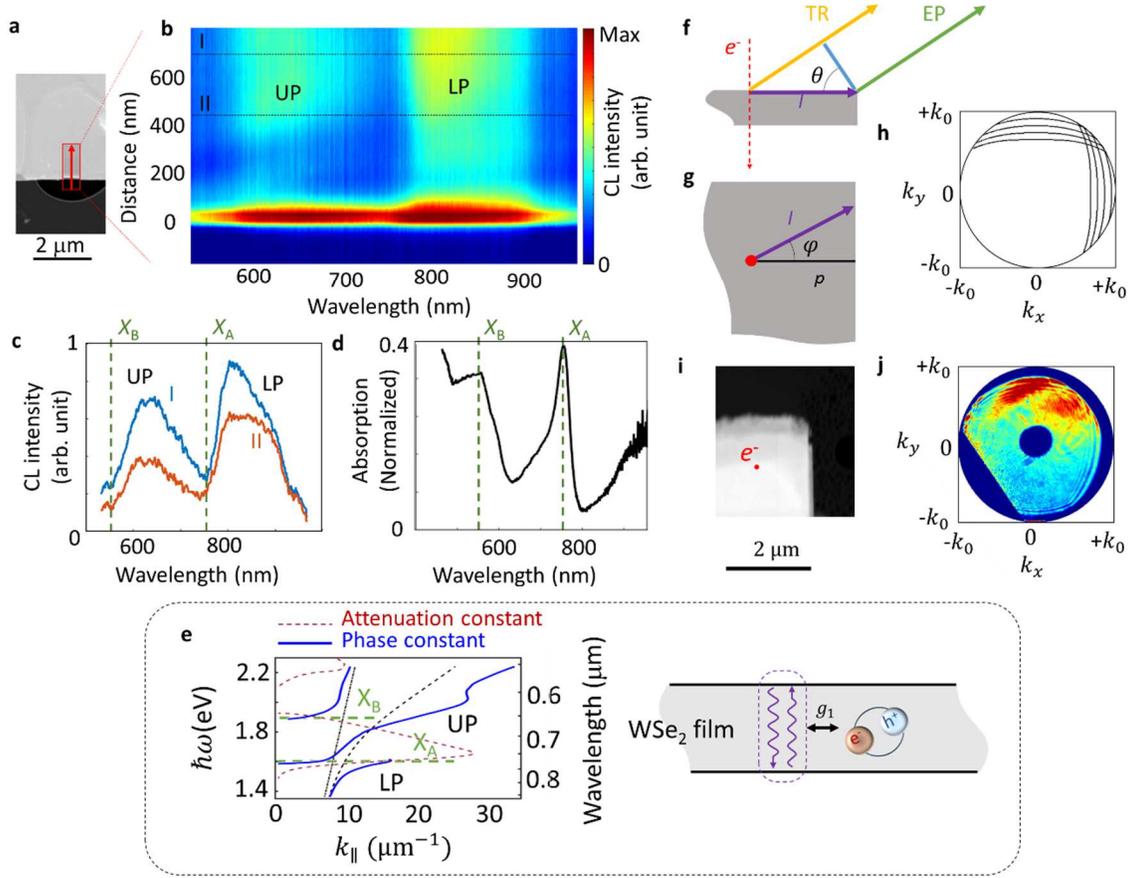

**Fig. S5. Exciton polaritons in the WSe₂ flake.** (a) SEM image of the sample, composed of a WSe₂ flake positioned on top of a holey carbon substrate. (b) CL spectra at different scanning positions of the electron beam. The sample is scanned along the scanning direction marked by the red arrow in (a) and integrated along the vertical direction. The formation of the LP and UP branches in the spectral regions marked by UP and LP, respectively, is apparent. (c) CL spectra at depicted electron impact positions marked by I and II in (b). (d) Absorption coefficient of the flake measured by using an all-optical dark-field microscopy/spectroscopy setup. (e) Propagation constant ($\beta - i\alpha$) of the exciton polaritons versus the photon energy ($\hbar\omega$), $\beta$ (the phase constant) and $\alpha$ (the attenuation constant) are shown by thick solid blue and dashed red curves, respectively. The optical line ($k_0 = \omega/c$) is indicated by a thin dashed-dotted black curve, and the dispersion of the photonic modes, due to the background permittivity $\varepsilon_{r,\infty} = 15$, is represented by a thin black dashed curve. X$_A$ and X$_B$ excitonic energies are marked by dashed green lines. (f) Radiation mechanisms are attributed to the transition radiation (TR; orange arrow), the excitation of exciton polaritons (purple arrow) and their scattering at the edge of the flake (green arrow). Shown is the lateral view. (g) Top view with the electron impact position shown by the red dot. The exciton polaritons propagating at an elevated angle $\varphi$ with respect to the normal to edge take the path length $l = p/\cos\varphi$ to reach the edge. (h) The position of the maxima of the interference maps modeled by assuming constructive interference between TR and the exciton polaritons at the far-field. The electron impact parameter is chosen to be $p = 1.6\ \mu m$ from both edges. (i) SEM image of the WSe₂ flake used to evaluate the TR/exciton-polariton interference hypothesis. (j) Angle-resolved CL pattern acquired at $\lambda = 800\ nm \pm 10\ nm$.

4. **Mutual coherence and simulated interference map**

Commonly, the visibility of the interference fringes determines the degree of coherence in all experiments involving a superposition of coherent waves. More precisely, the mechanism of decoherence can be tracked by observing how the visibility of the interference fringes is altered by systematically changing a control parameter, i.e., the distance between the individual components of a superposition. Thus, in our experiment, tracking the visibility of the interference fringes by altering the distance between the EDPHS and the sample allows us to determine the mutual coherence between the EDPHS and the sample radiation, and more prominently, observing the decoherence phenomena due to the interaction of the sample (and the EDPHS) excitations with the environment.

As mentioned in the main text, the angle-resolved radiation pattern can be divided into three domains (Fig. S6a and b): (i) in the region between $L = 22$ μm and 40 μm, the interference pattern is sharply altered, caused by the influence of the EDPHS radiation; (ii) in the range between $L = 40$ μm and 52 μm, the visibility of the interference fringes is completely diminished, (iii) followed by the backscattered waves caused by the TEM sample grid, the substrate, and possibly other flakes positioned on the holy carbon substrate, obvious in the range from $L = 52$ μm and 120 μm.

The observed far-field interference patterns, could be modeled as a superposition of the scattered waves from the flake, initiated by both the EDPHS and the flake (Fig. S7a). The contribution of the electron beam to the scattered waves are modeled by both transition radiation and the exciton polaritons scattered from the edge. Generalizing eq. (S1), for the case of a thinner flake, where the excited exciton polaritons could get scattered from both edges of the thin ribbon shown in Fig. 2a, we derive the electron induced CL as

$$\psi^{el}(\vec{r},\theta) = \frac{e^{-ik_0 r}}{4\pi r}\sin\theta\left\{1 + e^{ik_0\sin\theta l_1 - i(\beta-i\alpha)l_1}\sum_{n=0}^{+\infty}|T_2||R_2|^{2n}e^{-i(\beta-i\alpha)2nl_2}e^{i\varphi_{T_2}+i2n\varphi_{R_2}}\right.$$
$$\left. + e^{ik_0\sin\theta(l_2-l_1)-i(\beta-i\alpha)(l_2-l_1)}\sum_{n=1}^{+\infty}|T_2||R_2|^{2n}e^{-i\beta 2nl_2}e^{i\varphi_{T_2}+i2n\varphi_{R_2}}\right\}, \quad\text{(S3)}$$

where, the first term on the RHS, is due to the transition radiation. Each term in the first series is associated with the exciton polaritons with the phase constant $\beta$ that represent the scattering from the right edge of the flake after travelling *n* times through the flake and reflecting back from the left edge.

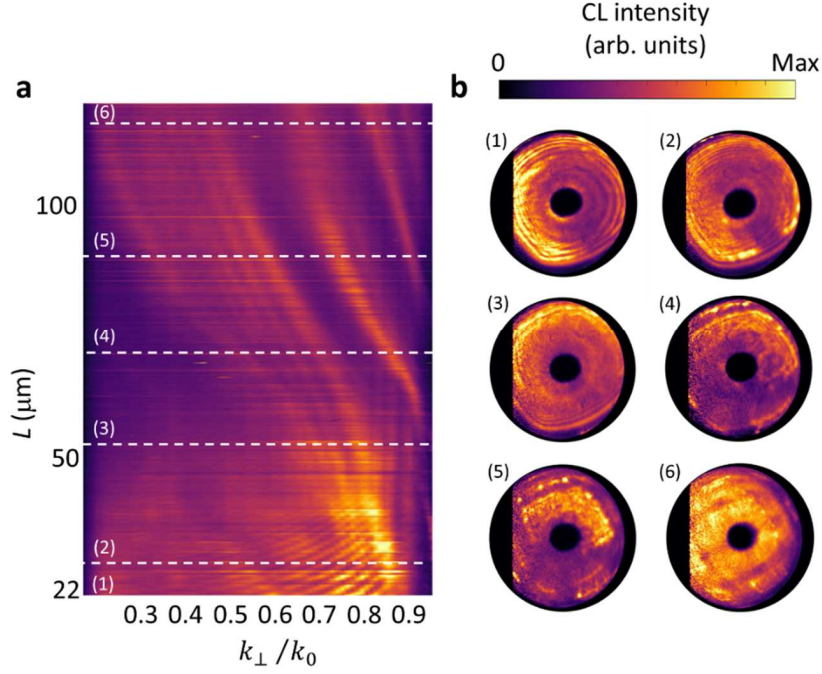

**Fig. S6. Mutual coherence and decoherence effects in EDPHS-sample interactions.** (a) Angle-resolved CL patterns at the wavelength of $\lambda = 800$ nm, versus the distance, integrated over the azimuthal angle range of 95° to 100°, and (b) at depicted distances $L$ between the EDPHS and the sample, demonstrating the modulation of the interferences patterns versus $L$.

The scattering from the edge of the flake is modeled by its transmission and reflection coefficients, $|T_2|e^{i\varphi_{T_2}}$ and $|R_2|e^{i\varphi_{R_2}}$ (Fig. S7b and c), respectively. The third term is due to the scattering of the exciton polaritons from the left side of the flake. The EDPHS scattered field can be modeled as

$$\psi^{\text{EDPHS}}(\vec{r},\theta) = \frac{e^{-ikr}}{4\pi r}\sin\theta \left\{ |T_1|e^{i\varphi_{T_1}} + e^{ik_0\sin\theta l_1}\sum_{n=1}^{+\infty}|T_2||R_2|^{2n-1}e^{-i(\beta-i\alpha)2nl_2}e^{i\varphi_{T_2}+i(2n-1)\varphi_{R_2}} \right.$$
$$\left. + e^{ik_0\sin\theta(l_2-l_1)-i(\beta-i\alpha)(l_2-l_1)}\sum_{n=0}^{+\infty}|T_2||R_2|^{2n}e^{-i(\beta-i\alpha)2nl_2}e^{i\varphi_{T_2}+i2n\varphi_{R_2}} \right\}, \quad (S4)$$

where the first term is due to the direct transmission through the flake. Its transmission and reflection coefficient are hence only frequency-dependent, due to the materials dispersion, and not dependent on the scattering angle $\theta$. Moreover, the $T_2$ and $R_2$ coefficients are only frequency-dependent, since the incident angle $\theta_1$ is related to the phase constant as $\theta_1 = \beta(\omega)/(\sqrt{\varepsilon_r(\omega)}k_0)$, where $\varepsilon_r(\omega)$ is the material permittivity. The total radiation is thus calculated as $|\psi|^2 = |\psi^{\text{el}} + e^{i\omega\tau}\psi^{\text{EDPHS}}|^2$, with the delay given by $\tau = L(v_{\text{el}}^{-1} - c^{-1})$. EDPHS radiation can also excite exciton polaritons, when the radiation is scattered from the edge of the flake, which is justified due to the transverse broadening of the EDPHS radiation (yellow circle in Fig. 2a). Thus, in first order ($n = 0$), only scattering from the right side of the flake is possible for the electron beam contribution, and the scattering from the left side is included in the EDPHS contribution. Figure S8 shows the comparison between two cases: (i) only scattering from the right side of the grid is included and terms up to $n = 1$ are considered, and (ii) scattering from the right side of the grid is also included, but still only terms up to the $n = 1$ are included. To calculate the transmission coefficients through the edge, the edges are modeled as an infinite interface. However, in reality the radiation will be in the form of a dipole oriented parallel to the edge of the flake, for flakes thinner than 100 nm, when the higher order multipolar excitations could be neglected. Thus, the dipole radiation pattern $\sin\theta$ is also retained for the scattered waves from the edge. The comparison between the cases, shows that indeed the major contribution is due to the scattering from the right side of the flake, though the scattering from the left side could cause lower frequency modulations along the $k_\parallel$ axis, in better agreement with the simulation and experimental results.

The angle-resolved maps acquired at fine steps of 100 nm within the range of $L = 0$ µm to 150 µm, required 36 hours of continuous measurements, where sample drift and beam instabilities might undermine a full agreement with the theoretical results, though apparently are minor issues due to our drift correction mechanism.

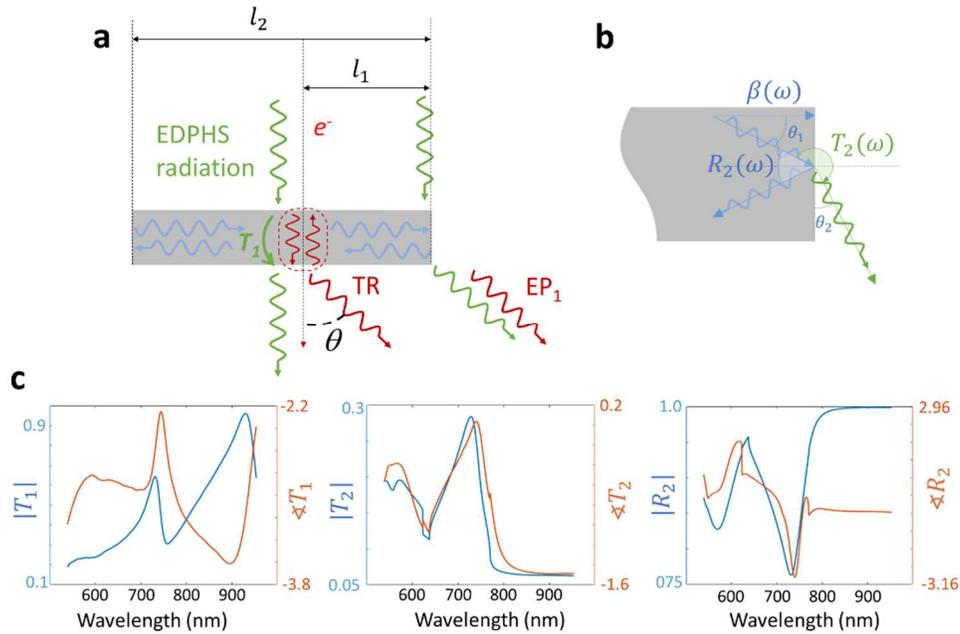

**Fig. S7. Optical pathways contributing to the overall CL pattern.** (a) Schematic of possible scattered beams, caused by the EDPHS radiation (green wavy arrows) and the electron beam (red wavy arrows). The blue wavy arrows show the excited guided mode of the thin film, to which both the electron beam and the EDPHS contribute. (b) Schematic of transmitted and reflected optical rays at the edge of the flake. (c) Analytically calculated amplitude and phase of the transmission and reflection coefficients $T_1$, $T_2$ and $R_2$.

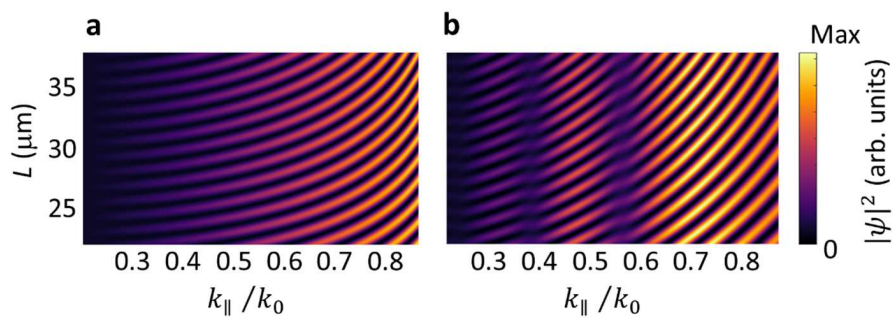

**Fig. S8. Analytically calculated interference fringes,** (a) including only radiation from the right side and (b) radiation from both the right and left sides of the flake.